\documentclass[a4paper,twoside]{article}

\usepackage[T1]{fontenc}

\usepackage{cite}
\usepackage{graphicx}
\usepackage{textcomp}
\usepackage{xcolor}
\usepackage{booktabs}
\usepackage{url}

\usepackage{epsfig}
\usepackage{subcaption}
\usepackage{calc}
\usepackage{amssymb}
\usepackage{amstext}
\usepackage{amsmath}
\usepackage{amsthm}
\usepackage{multicol}
\usepackage{pslatex}
\usepackage{apalike}
\usepackage{algorithm2e}
\usepackage[bottom]{footmisc}
\usepackage{SCITEPRESS}     

\begin{document}

\title{Got Ya! -- Sensors for Identity Management Specific \\Security Situational Awareness}

\author{\authorname{Daniela Pöhn\sup{1}\orcidAuthor{0000-0002-6373-3637}
 and Heiner Lüken\sup{1}}
\affiliation{\sup{1}University of the Bundeswehr Munich, RI CODE, Neubiberg, Germany}
\email{\{daniela.poehn, heiner.lueken\}@unibw.de}
}

\keywords{Security Situational Awareness, Cyber Situational Awareness, Security, Identity Management, OAuth.}

\abstract{Security situational awareness refers to identifying, mitigating, and preventing digital cyber threats by gathering information to understand the current situation. With awareness, the basis for decisions is present, particularly in complex situations. However, while logging can track the successful login into a system, it typically cannot determine if the login was performed by the user assigned to the account. An account takeover, for example, by a successful phishing attack, can be used as an entry into an organization's network. All identities within an organization are managed in an identity management system. Thereby, these systems are an interesting goal for malicious actors. Even within identity management systems, it is difficult to differentiate legitimate from malicious actions. We propose a security situational awareness approach specifically to identity management. 
We focus on protocol-specifics and identity-related sources in a general concept before providing the example of the protocol OAuth with a proof-of-concept implementation.}

\onecolumn \maketitle \normalsize \setcounter{footnote}{0} \vfill

\section{\uppercase{Introduction}}

The supply chain attack on SolarWinds' Orion platform~\cite{9604375} showed that identity and access management are crucial assets. The malicious actors took over Microsoft Active Directory (AD) instances and used the Federation Services (FS) extension to access other resources. Security mechanisms such as single sign-on (SSO) can be used against the target. According to \cite{9382367}, the attacker mimicked regular Security Assertion Markup Language (SAML) interactions for their purposes. However, not only malicious actors may target AD and SAML, but also other identity management (IdM)-related protocols, such as Open Authorization (OAuth) and OpenID Connect (OIDC). Detecting symptoms of attempts at an early stage may speed up the incident response process. This is difficult, as malicious activities are similar to regular interactions, and traditional security mechanisms may fail. However, this is not the only problem in this context. The variety of IdM protocols
~\cite{10148904} makes it even harder. Lastly, as we deal with digital identities, we must include humans. 

One potential way to improve the current situation is the application of security situational awareness tailored to IdM. Security or cyber situational awareness is an application of situational awareness in the cyber domain to perceive the environment (\emph{perception}), understand the current security situation (\emph{comprehension}), and project how the situation will evolve (\emph{projection}). Security situational awareness should provide the operators with a decision-making methodology in complex and sophisticated systems. Currently, several aspects in the field are being enhanced. According to \cite{10.1145/3384471}, humans are often not included, although they represent critical elements in this context. As shown above, it is also hard to differentiate legitimate from malicious user actions. Therefore, current security situational awareness approaches have to be adapted to identity management to include digital identities.

Consequently, we use the following research questions: 
What is the general outlook of security situational awareness that comprises identity management? 
What are suitable sensors and other sources for IdM-specific security situational awareness? 
What are the practical advantages of such an approach?

In order to increase the security of IdM, we propose (i.) a generic concept of security situational awareness specifically for IdM. Related to that, we focus on (ii.) sensors and other sources, as they are the first step for security situational awareness. We show  (iii.) the advantages of such an approach using the example of OAuth and discuss the implications. Therefore, this paper contributes (1) the derivation of the sensors and other sources related to IdM that are required for (2) a concept for IdM-specific security situational awareness and (3) the concept and proof-of-concept implementation for OAuth.

The remainder of the paper is as follows: We explain the background on identity management and situational awareness (see Section~\ref{sec:background}) and contrast it with related work (see Section~\ref{sec:sota}). Section~\ref{sec:concept} proposes the generic concept for security situational awareness for IdM before Section~\ref{sec:oauth} describes the example of OAuth. We summarize and discuss our approach in Section~\ref{sec:conclusion}.

\section{\uppercase{Background}}
\label{sec:background}


\subsection{Identity Management}
\label{sec:idm}

IdM comprises identification, authentication, authorization, and general management of users. In cross-organizational contexts, federated protocols like SAML, OAuth, and OIDC may be operated. As the identity providers can collect more data about the users in these protocols, self-sovereign identities (SSI) are currently introduced. The user has self-sovereign control over their data in a so-called wallet. OAuth and OIDC, the authentication protocol on top of OAuth, may be used in the SSI context by applying additional protocols, summarized as OpenID for Verifiable Credentials~\cite{oid4vp}.

\begin{figure}[h]
    \centering
    \includegraphics[width=\linewidth]{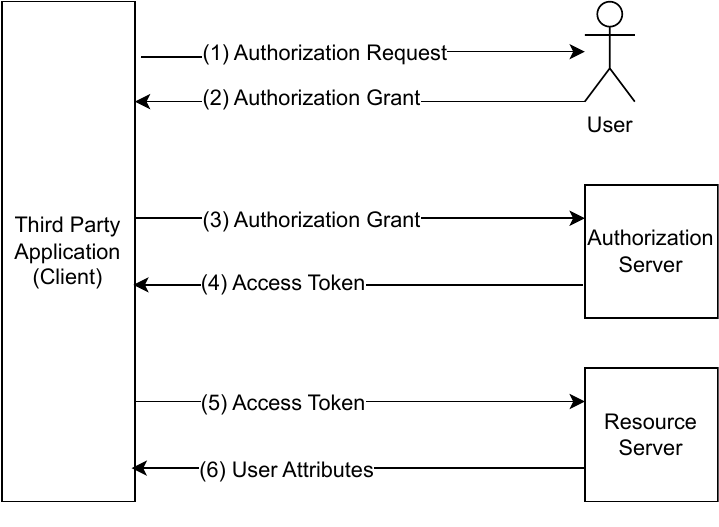}
    \caption{Generic workflow of the OAuth protocol.}
    \label{fig:oauth}
\end{figure}

\emph{OAuth~2.0}~\cite{rfc6749} is a protocol for authorization, i.e., OAuth permits users to share account information with third-party services and applications without providing them the credentials.

In Figure~\ref{fig:oauth}, the generic workflow is shown. The third-party applications request authorization from the resource owner (i.e., user; \emph{step 1}). The resource owner subsequently grants authorization (\emph{step 2}). The authorization grant is then forwarded to the authorization server (\emph{step 3}) that provides the third-party application with an access token (\emph{step 4}). The access token is then forwarded to the resource server (\emph{step 5}) to grant access (\emph{step 6}).

OAuth has specified different protocol flows, called grants, that enable authorization with variations of the workflow described above. However, not all flows are securely usable. The most common OAuth grant types are authorization code, proof key for code exchange (PKCE), client credentials, device code, and refresh token. In contrast, implicit flow and resource owner password credentials grant are insecure. Furthermore, OAuth-specific attack vectors are known~\cite{10.1145/2976749.2978385}, including OAuth access token abuse and theft~\cite{10.1145/3548606.3560692}, cross-site request forgery~\cite{10.1007/978-3-030-80825-9_2}, path confusion~\cite{10.1145/3627106.3627140}, and generally misconceptions~\cite{10.1145/2991079.2991105}. Security best practices are described in~\cite{rfc6819}. OAuth~2.1~\cite{ietf-oauth-v2-1} is an in-progress update (currently work-in-progress) that consolidates best practices and established extensions since OAuth~2.0 was published. Related to the grants, the common variants described above are included in OAuth~2.1.

\subsection{Situational Awareness}
\label{sec:csa}

Situational awareness describes the cognizance of entities in the environment (perception), understanding their meaning (comprehension), and the projection of their state in near future (projection) \cite{10.1177/154193128803200221}. Such a situational awareness is especially important in a military context. According to \cite{army}, ``situational understanding is the product of applying analysis and judgment to relevant information to determine the relationships within the situation.'' After determining the baseline, anomalies can be identified. These observations and the understanding of the anomalies in the context (situation assessment) are relevant for the projection. The three phases can be adapted for the cyber domain, speaking of security or cyber situational awareness. This means that security situational awareness is part of situational awareness which concerns the cyber environment.

According to \cite{FRANKE201418}, different sensors can be applied, including intrusion detection systems (IDSs), external information, and human intelligence. \cite{8073386} propose a taxonomy for cybersecurity situational awareness with a description of the scope, level, viewpoint, and decision making. Further, the authors divide data gathering into operational (i.e., antivirus, vulnerability scanner, penetration testing, network scanning, password cracking, firewall, and IDS) and strategic (i.e., asset listing, risk identification, surveys, incident response reports, audit findings, policy review, and news review). This lists however also shows that the data sources may have to be adapted to the network.

\section{\uppercase{Related Work}}
\label{sec:sota}

Various approaches target security situational awareness. \cite{10.1145/3407023.3407062} provide an overview of security situational awareness and explain contemporary challenges. The authors describe the toolset perspective through a taxonomy with several entries in the perception phase. Although most of them, like scans and log files, are relevant for IdM, other sensors and sources might be of value in our scenario. Similarly, surveys \cite{10216378,7917193,23052608} show several sources that mainly target networks. \cite{8073386} include password cracking. \cite{8753997} propose an approach of cyber situational awareness through social media, security news, and blog data mining, whereas \cite{8899406} focus on phishing attacks. Other methods to improve the comprehension phase include netflow visualization~\cite{10.1145/1029208.1029214} and graph-based approach~\cite{10.1007/978-3-031-44355-8_12}. These may be adaptable to IdM. Lastly, \cite{10.1145/3538969.3544414} provide an overview of offensive cyber operation automation tools that mostly focus on network security and, thus, might not be suitable for our purpose.

So far, no approach focuses on IdM, although digital identities and IdMS are the target of various attacks and current measures are not enough. When applying security situational awareness to IdM, the protocols and their attack characteristics must be considered. This is currently different in related work. In addition, other sensors specific to IdM, such as password leaks, data found in online sources, and the features used for risk-based authentication, may be included to provide a better picture.

\section{\uppercase{Identity Security Situational Awareness
}}
\label{sec:concept}

In this section, we describe the general concept with its layers, entities, and relations. We identify the layers of internal identity management, external  identity, threat, and detection. Within these layers, we focus on sensors and other data sources that can be used to fill the information required for a better picture. Based on the entities and relations, we identify patterns that can be applied to recognize anomalies.

\subsection{Internal Identity Management Layer}
\label{sec:idm-layer}

The internal identity management layer 
encompasses all data related to identity management, including digital identities and authentication.

\paragraph{Digital identities.} 
\label{sec:identity}

Users, data, applications, and devices all have an identity that is managed and belongs to an organization. To access data, applications, or devices, the user has to have the required permissions that may be provided to them, for example, due to roles or attributes. 
Following this, the management system can be used as a source of security situational awareness. The malicious and regular actor patterns do not differ significantly depending on the attack and exact system. A honeypot may be applicable for further information.


\paragraph{Authentication and authorization.} 
\label{sec:authnz}

Before using, for example, the service, the user has to be authenticated. The most common method is a password. However, other methods, including biometrics, may be applied. Depending on the permissions, the service decides on the authorization. Additional features, such as those applied by risk-based authentication, may provide additional value for both authentication and authorization. 
Again, the authentication methods are crucial for security, whereas unusual patterns may indicate a malicious actor.


\paragraph{Logged authentication actions.} 
\label{sec:logged}

When a user authenticates and uses a server, log files are written. Log files related to digital identities and identity management systems (IdMS) can be found at the system, IdMS, database, application, web server, and end-user application (such as an SSI wallet on a smartphone) level. The exact location depends on the actual application. The log files may include identifiers, such as session-related identifiers (IDs and tokens, among others), IP addresses, transaction IDs, device fingerprints and IDs, user IDs, and other identifiable data. The exact identifiers are subject to log format, log level, application, and policies, among others. 
Log files are an essential source for security situational awareness. However, identity-related attack patterns have to be considered.


\subsection{External Identity Layer}

It is crucial to note that the scope of identity management and digital identities extends beyond the internal organization. The external identity layer plays a significant role in this more complex context.

\paragraph{Cross-organizational identity management.} As summarized in Section~\ref{sec:background}, identity management is often not limited to a single organization. Different organizations may form a federation to use resources together. The identity management model (i.e., isolated, centralized, federated, or self-sovereign) has implications on the attack pattern that every organization can recognize, the impact, and the security incident response (SIR) process. 
Hence, another source for security situational awareness is the other organizations within a federation. 


\paragraph{Usage of external services.} 
\label{sec:external}

To make things more complex, users may use external services that are not part of a federation. 
For example, to submit a manuscript, the researcher creates an account with their name and work email address at a conference management software system, such as EasyChair. In social media, such as X (formerly Twitter) and LinkedIn, they may provide insights into their work. They may even reuse their work password, as shown by \cite{10.1145/1242572.1242661}. All these data may be used for malicious attempts. For example, password leaks allow malicious actors to stuff the credentials at other online services with the hope of taking over more accounts. If work-related data is included, these attacks may target the work organization. Hence, detecting password leakages is crucial, and these services, as described by \cite{10.1145/3319535.3354229}, should be included in the list of sources for security situational awareness. If the leak is not detected and added by typical leak services, it may be used by malicious actors. The sharing of personal information is even more complex to include. By sharing information, potential malicious actors may gather data that can be used for personalized attempts, such as social engineering attacks. Email spam filters, human sensors, and open-source intelligence honeypots like social media accounts may provide further input. Additionally, it might help to scan the Internet for identity-related information 
\cite{icissp23}.


\subsection{Threat Layer}

\cite{10.1145/3538969.3544430} differentiate IdMS, end-user, and service account in their attack taxonomy TaxIdMA. This differentiation is applied in the threat layer. 
All elements have an identity. Hence, all identities may become relevant during an attack. The threat layer describes the exposure of the network with its systems and users through vulnerabilities that may be within the software or users. 

Users, thereby, mainly refer to end-users. Nonetheless, IT personnel might be targeted later in the attack lifecycle. Typically, service accounts that have a vulnerability, such as a misconfiguration, are utilized during the attack lifecycle. The IdMS could be the final goal that can be reached by another vulnerability, such as a misconfiguration. Another way is to use another weakness, such as a common vulnerability enumeration (CVE) for a specific software version. Security events can be raised by security monitoring or humans.


\subsection{Detection Layer}
\label{sec:detection}

First, we derive the sensors before summarizing them. Similarly to the threat layer, we use the distinction between IdMS, end-user, and service account.


\paragraph{End-user.} 
Passwords can be guessed, for example, by spraying them on other services or brute-force attacks, including credential stuffing. Phishing is one of the primary initial access vector into an organization's network. However, variants of phishing involving social media, messenger, and other means are increasingly being used. 

Other threats may be human-in-the-middle (MitM), usage of malware, and session hijacking, to name a few examples. The malicious attackers may use a different pattern that is defined by the attack and the means. Relevant features may include the timing, session, device, IP address, username, and password. As mentioned above, these may depend on the actual system and attack. 
However, protocol specifics have to be taken into account, as outlined in Section~\ref{sec:background} for the protocol of OAuth.

\paragraph{Service user.} Regarding the attack lifecycle, service accounts might come next. These may have vulnerabilities related to software or configuration that allow the takeover by a malicious actor. In the next step, the malicious actor may try to find further vulnerabilities and systems. Following this, unusual behavior may hint at an attack. Thereby, we try to recognize anomalies that can be detected by host-based and network-based intrusion detection systems (HIDS/NIDS). Further research may improve the methods.

\paragraph{IdMS.} One of the main goals might be the IdMS, as seen with the SolarWinds' Orion hack. An IdMS can be targeted by taking over administration, deactivated, or unused accounts, or using software-related vulnerabilities. Depending on the IdMS, like AD, the actions performed by the malicious actor may be regular. This may increase the difficulty of detecting anomalies. Besides unusual actions, honeypot accounts and networks might be installed. However, such an approach is still up to future work.

\paragraph{Proposed sensors and context data.}
Table~\ref{tab:perception} summarizes the proposed sensors and context information. We suggest classical security countermeasures or sensors, some adaptations, and additional sensors. One example of adaptation is a honeypot applied to identity management to recognize attacks. Additionally, log file monitoring can help detect attempts. 

Regarding social engineering attacks and suspicious computer behavior, human-as-a-security-sensor (HaaSS) should be included (see, among others, Section~\ref{sec:external}). In addition to the sensor information, context information is required, according to \cite{10.1145/3465481.3470037}. Context information may include the location, users (see also the parameters used by risk-based authentication, which could be added to the log data), version of the OS, software, services, and vulnerabilities (see Section~\ref{sec:identity}
). Hence, a collection of this contextual data is needed. Preferably, it should be maintained constantly. Husák et al. propose the usage of NetFlow for passive monitoring and Nmap for active monitoring. 

In addition, the network plan, role concept (see Section~\ref{sec:identity}), and other documentation can help to notice undocumented changes. 

\begin{table*}
\centering
\caption{Input for perception as first phase of security situational awareness.}\label{tab:perception}
\begin{tabular}{p{3.9cm} p{3.9cm} p{3.9cm}}
\toprule
\textbf{Internal Input} & \textbf{Organizational Input} & \textbf{External input} \\ \midrule
Log files               & Role concept                  & Leak detection        \\
AV systems              & Policies                      & CERT warnings           \\
Firewalls               & IdM lifecycles                & Vulnerability reports   \\
IPS/IDS                     & IdMSecMan                     & Third-party reports     \\
OSINT/IdM honeypot                &          Network plan                 & OSINT/IdM honeypot                \\
NetFlow                 & Security concept              & Update information     \\
Network analyzer & Other documentation & Other external data\\
\multicolumn{3}{c}{OSINT framework and HaaSS}                                               \\
\bottomrule
\end{tabular}
\end{table*}

\subsection{Comprehension}

Comprehension describes the analysis, which requires methods and characteristics for IdM-related attacks. 
We outlined typical threats that have their pattern. These may have simple reasons, such as a device change since the smartphone was lost, or a malicious actor. These patterns can be observed in, for example, log files, which are the primary source of security situational awareness. Different identifiers might be included in the log files depending on the actual system and log level. The minimum characteristics required to recognize these attacks are summarized in Table~\ref{tab:char-overview}. More characteristics may provide a better picture if acceptable from a privacy perspective.

\begin{table*}
\centering
\caption{Overview of selected IdM-related attacks and their minimum characteristics.}
\label{tab:char-overview}
\begin{tabular}{ll}
\toprule
\textbf{Attack}     & \textbf{Characteristic}                                               \\ \midrule
Password stuffing   & Password (hashed), IP address(es)                                     \\
Wordlist            & Usernames, failures, IP address(es), maybe known input                \\
Credential stuffing & Usernames, IP address(es)                                             \\
Brute-force attack  & Usernames, failures, IP address(es)                                   \\
Session hijacking   & Different IP address and device/browser fingerprint                   \\
Phishing            & Email, login with different IP address and device/browser fingerprint \\
MitM & Different IP address and reuse of session-related data \\
Malware             & Traffic to external IP address(es), unusual behavior       \\ \bottomrule          
\end{tabular}
\end{table*}

\subsection{Projection}


As the focus is on IdM, we again use the IdM-specific information and the results from the comprehension phase. An overview of user accounts, their access and permissions on different systems, and a network plan help identify the extent of malicious actions. Information about previous and possible incidents provides further input. This can be visualized in dashboards. 

\section{\uppercase{Example of OAuth}}
\label{sec:oauth}

As outlined in Section~\ref{sec:sota}, OAuth in its current version~2.0 has some security drawbacks that are typically mitigated by applying the security best practices. For example, the \texttt{authlib} library has some inherent security properties enabled by default. In OAuth~2.1, changes are made to the security architecture of the framework, among other things~\cite{10148904}. The grants classified as insecure, the implicit grant and the resource owner password credentials grant, will be removed entirely from the framework and will, therefore, no longer be usable after the switch to version~2.1. We select the implicit and authorization code grants to see the differences between an insecure and a more secure grant. In addition, we use the threats described in~\cite{rfc6819} and general threats in web applications. We first set up a test environment to evaluate our exemplary implementation.

\subsection{Test Environment}

The test environment consists of a minimal OAuth setup 
with an authorization server, a resource server, and one client (i.e., all relevant entities) for each selected grant. Due to its wide usage, Python is chosen as a programming language in combination with the Flask framework and \texttt{authlib} library. A MySQL database is used to store persistent data. 

Based on Section~\ref{sec:detection}, the requests are logged in log files using the format of \texttt{TIMESTAMP - Source-IP: PORT - Request-Line - Header}. With these data, we should recognize all attacks in the test environment. Further features might be added to differentiate malicious from regular actions in the life environment. We select a rule-based approach as we have limited amount of data. In live environments, isolation forest and hidden Markov models might be better suited. The rules are created based on known attacks, such as the reuse and replay of tokens, cross-site scripting, and MitM attacks, while applying OAuth protocol specifics.

In order to automate user actions, i.e., regular and malicious actions, we use scripts. The malicious requests can be started using this particular script. The log files are used as input for the anomaly detection. When malicious requests are recognized by this anomaly detection, they are then displayed in the graphical user interface (GUI) for the IT personnel.

\subsection{Practical Example}

We apply the typical OAuth workflow, as described in Section~\ref{sec:background}, in our example. 

\paragraph{Creating a client.}
As a prerequisite, a client has to be created. A legitimate user is authenticated and gets redirected to their user profile. After clicking the `create client' button, a client is registered in the user's name. The client's metadata is stored in the MySQL database. Then, the client can initiate an OAuth grant. By pressing the `Login with OAuth' button, the URL is redirected to the server. The request is also logged in the log file. If the user has no current session, they are redirected to the login site.

\paragraph{Creating a session.}
Next, a session is created. As soon as the user is authenticated to the server, the server must agree to client-side access to their data. When the user has authorized access to their data through the client application (step 1), an \texttt{authorization\_code} is created by the server and sent to the client (step 2). The client now creates the \texttt{client\_code} and sends it together with the received \texttt{authorization\_code} as a post request (step 3). The server then creates a token, stores it in the MySQL database, and sends it to the client as a header entry (step 4). The client immediately redirects the user to the user profile page. The user name and email are displayed on the user profile. The client itself does not have access to this data because it is stored in the server's database. To obtain this data, the client must send the previously received \texttt{access\_token} to the server API in the form of a GET request as the authorization header (step 5).

\begin{figure}[h]
    \centering
    \includegraphics[width=\linewidth]{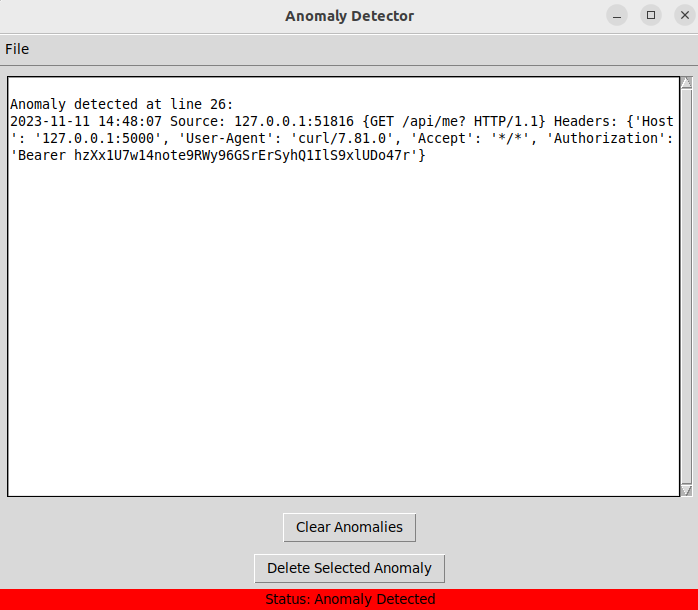}
    \caption{Security personnel GUI showing a malicious OAuth request.}
    \label{fig:tool}
\end{figure}

\paragraph{Detecting an anomaly.}
The testing of the OAuth test environment resulted in the creation of two tokens: \texttt{kvGWM72HDhLmatAoIiIxwgUbIhY92elmFs9 DkKKlht} for the authorization code grant and \texttt{XfKybKsgXU61HuBy1Kpy8Dy85GPj3TwKbpQSlRJnAd} for the implicit grant. Every client's access to protected resources results in the creation of a new token. The server saves all incoming requests. The anomaly detection application reads the created log file \texttt{header\_logs.log} in a loop. The anomaly detection application recognizes and displays tokens used multiple times as anomalies. If the malicious actor (in the test environment, it is our script) has gained the token, for example, via a human-in-the-middle attack, they may reuse it to request access to resources. If the malicious actor uses the token of the authorization code grant to request protected resources, this request results in an anomaly and is displayed in the security personnel GUI, as shown in Figure~\ref{fig:tool}. As a second aspect, the user-agent \texttt{curl} would also result in an anomaly.

The number of features that can be used for the rules depends on the information logged. The accuracy also depends on the details and number of rules used. Furthermore, other sensors add valuable data, such as leaked passwords on other websites, and OAuth-specific rules should not be the only source.




\section{\uppercase{Conclusion and Outlook}}
\label{sec:conclusion}

As identity and access management are crucial assets that malicious actors target, the security measures have to adjust and consider the specifications of identity management and its protocols. We proposed a generic security situational awareness approach specific to IdM to improve awareness in this field. Based on related work, we noticed that risk-based authentication uses several features that can be applied as sensors. Next, we identified several sensors, other sources, and attack patterns. Then, we practically showed the test environment and proof-of-concept implementation for OAuth.

The proposed concept is the first one specific to identity management. It comprises various sources and 
may provide the security personnel with further insights into the security of their infrastructure. However, as it mainly concentrates on sensors, other sources, and attack patterns, it is only the first step towards security situational awareness. As identity management is directly linked to the handling of personally identifiable information, we were not allowed to use a practical example with (pseudonymized) real world logs. Instead, a test environment for OAuth was chosen that will be extended in future work. Following this, the efficiency of the approach was not evaluated, since the detection was done in real time within the test environment. To the best of our knowledge, this is the first IdM-specific approach. By switching the focus to identities, insights might be gathered, as shown with the example of OAuth.

However, it is the first step, and the implementation has to be extended in future work. In addition, we want to integrate external sensors and evaluate the approach with actual data. Since identity management is often cross-organizational, we finally want to investigate in cross-organizational situational awareness.


\bibliographystyle{apalike}
{\small
\bibliography{situational-awareness}}

\begin{thebibliography}{}

\bibitem[Benolli et~al., 2021]{10.1007/978-3-030-80825-9_2}
Benolli, M., Mirheidari, S.~A., Arshad, E., and Crispo, B. (2021).
\newblock {The Full Gamut of an Attack: An Empirical Analysis of OAuth CSRF in
  the Wild}.
\newblock In Bilge, L., Cavallaro, L., Pellegrino, G., and Neves, N., editors,
  {\em Detection of Intrusions and Malware, and Vulnerability Assessment},
  pages 21--41, Cham. Springer International Publishing.

\bibitem[Endsley, 1988]{10.1177/154193128803200221}
Endsley, M.~R. (1988).
\newblock {Design and Evaluation for Situation Awareness Enhancement}.
\newblock {\em Proceedings of the Human Factors Society Annual Meeting},
  32(2):97--101.

\bibitem[Evesti et~al., 2017]{8073386}
Evesti, A., Kanstrén, T., and Frantti, T. (2017).
\newblock Cybersecurity situational awareness taxonomy.
\newblock In {\em International Conference On Cyber Situational Awareness, Data
  Analytics And Assessment (Cyber SA)}, pages 1--8.

\bibitem[Fett et~al., 2016]{10.1145/2976749.2978385}
Fett, D., K\"{u}sters, R., and Schmitz, G. (2016).
\newblock {A Comprehensive Formal Security Analysis of OAuth 2.0}.
\newblock In {\em Proceedings of the 2016 ACM SIGSAC Conference on Computer and
  Communications Security (CCS)}, page 1204–1215.

\bibitem[Florencio and Herley, 2007]{10.1145/1242572.1242661}
Florencio, D. and Herley, C. (2007).
\newblock A large-scale study of web password habits.
\newblock In {\em Proceedings of the 16th International Conference on World
  Wide Web (WWW)}, page 657–666.

\bibitem[Franke and Brynielsson, 2014]{FRANKE201418}
Franke, U. and Brynielsson, J. (2014).
\newblock Cyber situational awareness – a systematic review of the
  literature.
\newblock {\em Computers \& Security}, 46:18--31.

\bibitem[Gutzwiller et~al., 2020]{10.1145/3384471}
Gutzwiller, R., Dykstra, J., and Payne, B. (2020).
\newblock {Gaps and Opportunities in Situational Awareness for Cybersecurity}.
\newblock {\em Digital Threats}, 1(3).

\bibitem[Hardt, 2012]{rfc6749}
Hardt, D. (2012).
\newblock {The OAuth 2.0 Authorization Framework}.
\newblock RFC 6749.

\bibitem[Hardt et~al., 2024]{ietf-oauth-v2-1}
Hardt, D., Parecki, A., and Lodderstedt, T. (2024).
\newblock {The OAuth 2.1 Authorization Framework}.
\newblock Internet-Draft draft-ietf-oauth-v2-1-10, Internet Engineering Task
  Force.
\newblock Work in Progress.

\bibitem[{Headquarters, Department of the Army}, 2021]{army}
{Headquarters, Department of the Army} (2021).
\newblock {Advanced Situational Awareness}.
\newblock Technical report.

\bibitem[Hus\'{a}k et~al., 2020]{10.1145/3407023.3407062}
Hus\'{a}k, M., Jirs\'{\i}k, T., and Yang, S.~J. (2020).
\newblock {SoK}: contemporary issues and challenges to enable cyber situational
  awareness for network security.
\newblock In {\em Proceedings of the 15th International Conference on
  Availability, Reliability and Security (ARES)}.

\bibitem[Hus{\'a}k et~al., 2023]{10.1007/978-3-031-44355-8_12}
Hus{\'a}k, M., Khoury, J., Klisura, D., and Bou-Harb, E. (2023).
\newblock {On the Provision of Network-Wide Cyber Situational Awareness via
  Graph-Based Analytics}.
\newblock In Collet, P., Gardashova, L., El~Zant, S., and Abdulkarimova, U.,
  editors, {\em Complex Computational Ecosystems}, pages 167--179, Cham.
  Springer Nature Switzerland.

\bibitem[Hus\'{a}k et~al., 2021]{10.1145/3465481.3470037}
Hus\'{a}k, M., La\v{s}tovi\v{c}ka, M., and Tovar\v{n}\'{a}k, D. (2021).
\newblock {System for Continuous Collection of Contextual Information for
  Network Security Management and Incident Handling}.
\newblock In {\em Proceedings of the 16th International Conference on
  Availability, Reliability and Security (ARES)}.

\bibitem[Innocenti et~al., 2023]{10.1145/3627106.3627140}
Innocenti, T., Golinelli, M., Onarlioglu, K., Mirheidari, A., Crispo, B., and
  Kirda, E. (2023).
\newblock {OAuth 2.0 Redirect URI Validation Falls Short, Literally}.
\newblock In {\em Proceedings of the 39th Annual Computer Security Applications
  Conference (ACSAC)}, page 256–267.

\bibitem[Jannett et~al., 2022]{10.1145/3548606.3560692}
Jannett, L., Mladenov, V., Mainka, C., and Schwenk, J. (2022).
\newblock Distinct: Identity theft using in-browser communications in
  dual-window single sign-on.
\newblock In {\em Proceedings of the 2022 ACM SIGSAC Conference on Computer and
  Communications Security (CCS)}, page 1553–1567.

\bibitem[Kom\'{a}rkov\'{a} et~al., 2018]{10.1145/3230833.3232798}
Kom\'{a}rkov\'{a}, J., Hus\'{a}k, M., La\v{s}tovi\v{c}ka, M., and
  Tovar\v{n}\'{a}k, D. (2018).
\newblock {CRUSOE: Data Model for Cyber Situational Awareness}.
\newblock In {\em Proceedings of the 13th International Conference on
  Availability, Reliability and Security (ARES)}.

\bibitem[Legg and Blackman, 2019]{8899406}
Legg, P. and Blackman, T. (2019).
\newblock {Tools and Techniques for Improving Cyber Situational Awareness of
  Targeted Phishing Attacks}.
\newblock In {\em International Conference on Cyber Situational Awareness, Data
  Analytics And Assessment (Cyber SA)}, pages 1--4.

\bibitem[Li et~al., 2019]{10.1145/3319535.3354229}
Li, L., Pal, B., Ali, J., Sullivan, N., Chatterjee, R., and Ristenpart, T.
  (2019).
\newblock {Protocols for Checking Compromised Credentials}.
\newblock In {\em Proceedings of the 2019 ACM SIGSAC Conference on Computer and
  Communications Security (CCS)}, page 1387–1403.

\bibitem[Lodderstedt et~al., 2013]{rfc6819}
Lodderstedt, T., McGloin, M., and Hunt, P. (2013).
\newblock {OAuth 2.0 Threat Model and Security Considerations}.
\newblock RFC 6819.

\bibitem[Nour et~al., 2023]{10216378}
Nour, B., Pourzandi, M., and Debbabi, M. (2023).
\newblock {A Survey on Threat Hunting in Enterprise Networks}.
\newblock {\em IEEE Communications Surveys \& Tutorials}, 25(4):2299--2324.

\bibitem[{OpenID Foundation}, 2024]{oid4vp}
{OpenID Foundation} (2024).
\newblock {OpenID for Verifiable Credentials - Overview}.
\newblock \url{https://openid.net/sg/openid4vc/}.

\bibitem[Peisert et~al., 2021]{9382367}
Peisert, S., Schneier, B., Okhravi, H., Massacci, F., Benzel, T., Landwehr, C.,
  Mannan, M., Mirkovic, J., Prakash, A., and Michael, J.~B. (2021).
\newblock {Perspectives on the SolarWinds Incident}.
\newblock {\em IEEE Security \& Privacy}, 19(2):7--13.

\bibitem[P\"{o}hn and Hommel, 2022]{10.1145/3538969.3544430}
P\"{o}hn, D. and Hommel, W. (2022).
\newblock {TaxIdMA: Towards a Taxonomy for Attacks related to Identities}.
\newblock In {\em Proceedings of the 17th International Conference on
  Availability, Reliability and Security (ARES)}.

\bibitem[Pöhn and Hommel, 2023]{10148904}
Pöhn, D. and Hommel, W. (2023).
\newblock {New Directions and Challenges within Identity and Access
  Management}.
\newblock {\em IEEE Communications Standards Magazine}, 7(2):84--90.

\bibitem[Rodriguez and Okamura, 2019]{8753997}
Rodriguez, A. and Okamura, K. (2019).
\newblock {Generating Real Time Cyber Situational Awareness Information Through
  Social Media Data Mining}.
\newblock In {\em 43rd Annual Computer Software and Applications Conference
  (COMPSAC)}, volume~2, pages 502--507.

\bibitem[Sterle and Bhunia, 2021]{9604375}
Sterle, L. and Bhunia, S. (2021).
\newblock {On SolarWinds Orion Platform Security Breach}.
\newblock In {\em {SmartWorld, Ubiquitous Intelligence \& Computing, Advanced
  \& Trusted Computing, Scalable Computing \& Communications, Internet of
  People and Smart City Innovation (SmartWorld/SCALCOM/UIC/ATC/IOP/SCI)}},
  pages 636--641.

\bibitem[Tianfield, 2016]{7917193}
Tianfield, H. (2016).
\newblock {Cyber Security Situational Awareness}.
\newblock In {\em International Conference on Internet of Things (iThings) and
  IEEE Green Computing and Communications (GreenCom) and IEEE Cyber, Physical
  and Social Computing (CPSCom) and IEEE Smart Data (SmartData)}, pages
  782--787.

\bibitem[Walkow and Pöhn, 2023]{icissp23}
Walkow, M. and Pöhn, D. (2023).
\newblock {Systematically Searching for Identity-Related Information in the
  Internet with OSINT Tools}.
\newblock In {\em Proceedings of the 9th International Conference on
  Information Systems Security and Privacy (ICISSP)}, pages 402--409.

\bibitem[Wang et~al., 2016]{10.1145/2991079.2991105}
Wang, H., Zhang, Y., Li, J., and Gu, D. (2016).
\newblock {The Achilles heel of OAuth: a multi-platform study of OAuth-based
  authentication}.
\newblock In {\em Proceedings of the 32nd Annual Conference on Computer
  Security Applications (ACSAC)}, page 167–176.

\bibitem[Yin et~al., 2004]{10.1145/1029208.1029214}
Yin, X., Yurcik, W., Treaster, M., Li, Y., and Lakkaraju, K. (2004).
\newblock {VisFlowConnect}: netflow visualizations of link relationships for
  security situational awareness.
\newblock In {\em Proceedings of the 2004 ACM Workshop on Visualization and
  Data Mining for Computer Security (VizSEC/DMSEC)}, page 26–34.

\bibitem[Zhang et~al., 2023]{23052608}
Zhang, J., Feng, H., Liu, B., and Zhao, D. (2023).
\newblock {Survey of Technology in Network Security Situation Awareness}.
\newblock {\em Sensors}, 23(5).

\bibitem[Zurowski et~al., 2022]{10.1145/3538969.3544414}
Zurowski, S., Lord, G., and Baggili, I. (2022).
\newblock {A Quantitative Analysis of Offensive Cyber Operation (OCO)
  Automation Tools}.
\newblock In {\em Proceedings of the 17th International Conference on
  Availability, Reliability and Security (ARES)}.

\end{thebibliography}

\end{document}